# Experimental demonstration of quantum key distribution without monitoring of the signal disturbance

Shuang Wang,[1] Zhen-Qiang Yin,[1] Wei Chen,[1, *] De-Yong He,[1] Xiao-Tian Song,[1] Hong-Wei Li,[1] Li-Jun Zhang,[1] Zheng Zhou,[1] Guang-Can Guo,[1] and Zheng-Fu Han[1, †]

[1] *Key Laboratory of Quantum Information, University of Science and Technology of China, Hefei 230026, China and Synergetic Innovation Center of Quantum Information & Quantum Physics, University of Science and Technology of China, Hefei, Anhui 230026, China*

(Dated: April 25, 2015)

Quantum key distribution (QKD) enables two distant users, Alice and Bob, to share secret keys. In existing QKD protocols [1-3], an eavesdropper's intervention will inevitably disturb the quantum signals; thus, Alice and Bob must monitor the signal disturbance to place a bound on the potential information leakage. However, T. Sasaki et al. proposed a quite different protocol [4], named round-robin differential phase shift (RRDPS), in which the amount of eavesdropped information is bounded without monitoring the signal disturbance. Here, we present the first active implementation of the RRDPS protocol. In our experiment, Alice prepares packets of pulses, with each packet being a train with 65 pulses, and the global phase of each packet is randomized. Bob uses a 1-GHz, 1-64-bit actively controlled variable-delay interferometer to realize random switching of the different delays. Benefiting from the large pulse number of each packet, the high stability and low insertion loss of the interferometer, the system can distribute secret key over a distance of 90 km. Our results confirm the feasibility of performing practical QKD with this novel protocol.

Since the first QKD protocol (BB84) [1] was proposed in 1984, great progress have been made in QKD experiments [5-18]. A major step in QKD is transmitting quantum signals between Alice and Bob. Because of Heisenberg's uncertainty principle, an eavesdropper's intervention with the quantum signals transmitted between two users inevitably disturbs the quantum signals. Therefore, all existing QKD protocols must monitor the signal disturbance to place a bound on the potential information leakage [3]. However, T. Sasaki et al. proposed a quite different protocol [4], named round-robin differential phase shift (RRDPS), in which the amount of eavesdropped information is bounded without monitoring the signal disturbance. Here, we present the first active implementation of the RRDPS protocol.

Before proceeding to our experiment, it is beneficial to introduce the workflow of the RRDPS protocol and explain why this protocol is quite different from the previous QKD protocols. For simplicity, we first consider the case in which Alice is equipped with a perfect single-photon source. The workflow is as follows: (I) Alice generates an L-bit random sequence $s_1 s_2 \ldots s_L \in \{0,1\}^{\otimes L}$ and encodes her single-photon quantum signal as a superposition of time bins:

$$|\Psi\rangle = \frac{1}{\sqrt{L}} \sum_{k=1}^{L} (-1)^{s_k} |k\rangle, \quad (1)$$

in which $|k\rangle$ ($1 \leq k \leq L$) represents the case in which the single photon is in the $kth$ time-bin. (II) With a variable-delay interferometer, Bob randomly sets the pulse delay value ($1 \leq r \leq L-1$) and attempts to observe the relative phase between $|k\rangle$ and $|k+r\rangle$ ($k+r \leq L$). If the relative phase between $|k\rangle$ and $|k+r\rangle$ is obtained, the value of $k$ and $r$ will be declared to Alice via an authenticated public channel. Meanwhile Bob records his sifted key bit as 0 (1) if the corresponding relative phase is 0 ($\pi$). (III) Alice computes $s = s_k \oplus s_{k+r}$ as her sifted key bit, where $\oplus$ denotes summation modulo 2. (IV) Repeating steps (I)−(III), Alice and Bob can accumulate sufficient sifted key bits. Then, the final secure key bits can be generated via error correction and privacy amplification. Intuitively, there is only one photon, and the delay value $r$ is unknown to Eve before sifting; thus, the chance that Eve can learn the phase difference between $|k\rangle$ and $|k+r\rangle$ is small when $L$ is large. The larger $L$ is, the smaller the chance of eavesdropping is. In Ref. [4], it is proved that Eve's information about the sifted key bits is bounded by $h(1/(L-1))$, where $h(x) = -x \log_2 x - (1-x) \log_2 (1-x)$ is the Shannon entropy function. Here, it is clear that Eve's information is estimated without using the bit error rate; thus, monitoring of the signal disturbance is not necessary for privacy amplification. The physics behind the security mechanism is information causality [19], which forbids Eve from learning key bits.

The above arguments represent an ideal model only in theory because the assumed single-photon source is not available at present. In real-life situations, Alice must prepare packets of L weak coherent pulses instead of single-photon pulses. Assuming that the average photon number per pulse is $\mu$, the probability that there are more than $v_{th}$ photons in one packet is given by

$$e_{src} \triangleq Pr(v \geq v_{th}) = 1 - \sum_{n=0}^{v_{th}} \frac{e^{-L\mu}(L\mu)^n}{n!}, \quad (2)$$

where, $v$ represents the photon number in this packet. When the global phase of each packet is randomized, the final asymptotic secret key rate per packet can be given by [4]

$$R = Q(1 - fh(e_b)) - e_{src} - (Q - e_{src})h(\frac{v_{th}}{L-1}), \quad (3)$$

in which $f$ is the efficiency of error correction, $Q$ is the yield per packet, and $e_b$ represents the error rate of the sifted key bits. For a set of given experimental parameters $L$, $\mu$, $Q$ and $e_b$, one should optimize $v_{th}$ to maximize the secret key bit rate $R$.

For the finite-sized sifted key case with a failure probability $3 \times 2^{-s}$, the secret key rate can be corrected as follows [4]:

$$R_{ft} = [N(1 - fh(e_b)) - Nr_1 - N(1 - r_1)h(r_2) - s]/N_{em}, \quad (4)$$

where $N_{em}$ is the number of emitted packets, $N$ is the corresponding length of the sifted key bits, $r_1 = p_1 + \sqrt{(2ln2)p_1(1-p_1)s/N}$, $r_2 = p_2 + \sqrt{(2ln2)p_2(1-p_2)s/(N(1-r_1))}$, and $p_1 = e_{src}N_{em}/N$, and $p_2 = v_{th}/(L-1)$.

From (3) and (4), it is quite clear that the larger $L$ is, the higher the secret key rate we obtain. When $L$ is sufficiently large, RRDPS may outperform the existing protocols, such as BB84. However, it is not easy to implement the RRDPS protocol with large $L$. The main difficulty lies in implementing a variable-delay interferometer with high visibility and good stability. To overcome this challenge, we propose four typical schemes to implement such a variable-delay interferometer (for three of the schemes, see the Methods for details), and we use the optimal Faraday-Michelson scheme to experimentally realize RRDPS QKD with $L = 65$.

The experimental setup is summarized in Fig. 1. At Alice's site, a continuous wave (CW)

laser, whose central wavelength is 1550.12 nm, is first modulated into a pulse train by the subsequent intensity modulator ($IM_1$). The pulse in the train has a temporal width of 96 ps and a repetition rate of 1 GHz. Next, the second intensity modulator ($IM_2$) chops this pulse train into packets of 65 pulses. By randomly modulating $\{-\frac{\pi}{2}, \frac{\pi}{2}\}$ on each pulse, the key data are encoded on the packet using the first phase modulator ($PM_1$). The second phase modulator ($PM_2$) is added to randomize the global phase of each packet [20]. Additionally, the following variable attenuator (VA) attenuates the average photon number per pulse to the optimal value, which is obtained from numerical simulations (see Methods for details). These coded packets are then transmitted to Bob via a standard telecom fiber channel. At Bob's site, a three-port circulator (CIR) is set before his 1-GHz, 1-64-bit variable-delay Faraday-Michelson interferometer (FMI), which makes the $k$th pulse interfere with the $(k+r)$th pulse in the same packet; here, $r$ is the current delay of the interferometer, which ranges from 1 to 64. The two outputs of the FMI are connected to a double-channel single-photon detector (SPD). Both the sync signals and outputs of the SPD are sent to the time-to-digital convertor (TDC). Once the TDC records a click event in the interference area, Bob and Alice can share one sifted key bit.

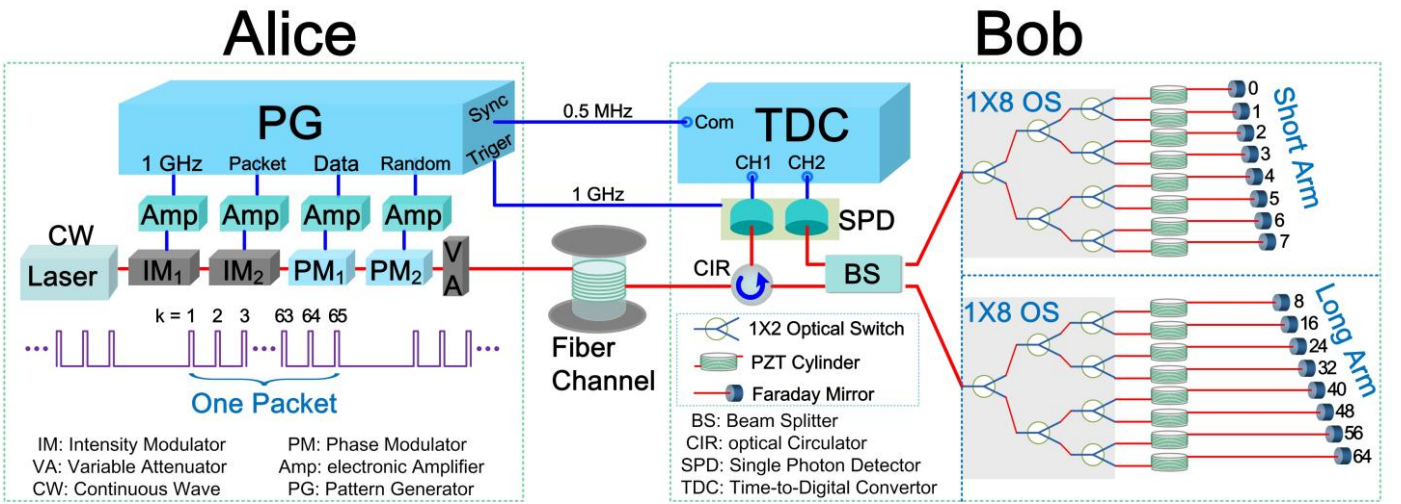

**FIG. 1.** Experimental setup of the RRDPS QKD system. Using $IM_1$ and $IM_2$, Alice prepares packets that are composed of a train of 65 pulses separated by intervals of 1 ns. These pulses are encoded with a randomly modulated phase $\{-\frac{\pi}{2}, \frac{\pi}{2}\}$. $PM_2$ is added to randomize the global phase of each packet. Bob uses a 1-GHz, 1-64-bit variable-delay interferometer and a double-channel SPD to observe the relative phase between the $k$th and $(k+r)$th pulses, where $r$ is the random delay of the variable-delay interferometer that Bob sets for the received packet, $1 \leq r \leq 64$, and $1 \leq k \leq 65 - r$. The variable-delay interferometer is based on the structure of the FMI, which is insensitive to polarization variation. Once the delay $r = x - y$ is set by controlling two $1 \times 8$ OSs, the interferometer has a long arm with delay $x \in \{8, 16, 24, 32, 40, 48, 56, 64\}$ and a short arm with delay $y \in$

$\{0, 1, 2, 3, 4, 5, 6, 7\}$. In each delay, a PZT cylinder wrapped with an optical fiber is utilized to obtain better precision and to compensate for the phase shift. The gated double-channel SPD is based on InGaAs avalanche photodiodes and uses the sine-wave filtering method. The maximum transmission distance of the fiber channel in the experiment is 90 km.

The variable-delay interferometer is the core part of Bob's setup. Here, we use the FMI [21] as the basic structure. One FMI is composed of one 50/50 beam splitter (BS) and two Faraday mirrors (FMs), which make the interferometer insensitive to polarization variation. The stability of this fixed-delay FMI has been demonstrated in several previous QKD experiments [7,8,16,21,22]. To realize the variable-delay interferometer, each of the two arms of the conventional FMI were replaced with one $1 \times 8$ optical switch (OS) and eight FMs with different fiber delays. Each $1 \times 8$ OS is composed of seven NanoSpeed $1 \times 2$ OSs (from Agiltron Inc.). The arm with the $\{0,1,2,3,4,5,6,7\}$ delays is called the short arm, whereas that with $\{8,16,24,32,40,48,56,64\}$ delays is called the long arm. One $1 \times 8$ OS independently chooses the $x$ delay in the long arm, and the other $1 \times 8$ OS chooses the $y$ delay in the short arm. Once the $x$ and $y$ delays are decided, the total delay of the interferometer is $x - y = r$. In total, there are 64 combinations of $x$ and $y$ delays. See Fig. 2a for the design that transforms the interferometer into a 1-GHz, 1-64-bit variable-delay interferometer.

One of the main challenges for realizing this variable-delay interferometer is to obtain each delay with high precision. In conventional DPS QKD experiments, the repetition rate of the pulse train can be finely tuned to match the fixed-delay interferometer because there is only one delay value. In contrast, in the variable-delay interferometer, there are too many delays to match. In this 1-GHz, 1-64-bit variable-delay FMI, the final precision of each delay is only approximately 10 μm, despite significant efforts made to improve the precision. Furthermore, a piezoelectric transducer (PZT) cylinder wrapped with an optical fiber was utilized to obtain better precision. Each delay in the long and short arms has one PZT cylinder whose stretching coefficient is approximately 0.08 μm/(V · m), and the length of the optical fiber that is wrapped around the cylinder is approximately 2 meters. In addition, the phase shift in each delay was compensated by carefully adjusting the voltage on each PZT cylinder.

The average insertion loss (IL) of the 1-GHz, 1-64-bit variable-delay interferometer is 5.60 dB. Compared with our previous fixed-delay interferometer, the increment of IL mainly comes from the round-trip IL of three $1 \times 2$ OSs. Each IL value of 64 delays is shown in Fig. 2b. The maximum IL in the long arm is at $x = 48$, and that in the short arm is at $y = 2$; thus, the maximum IL of the variable-delay interferometer, 6.04 dB, occurs at a delay value of 46. Similarly, the minimum IL, 5.30 dB, occurs at a delay value of 5 ($x = 8, y = 3$). In the worst case, the discrepancy between the long arm and short arm is only approximately 0.37 dB, which has a very small impact on the visibility.

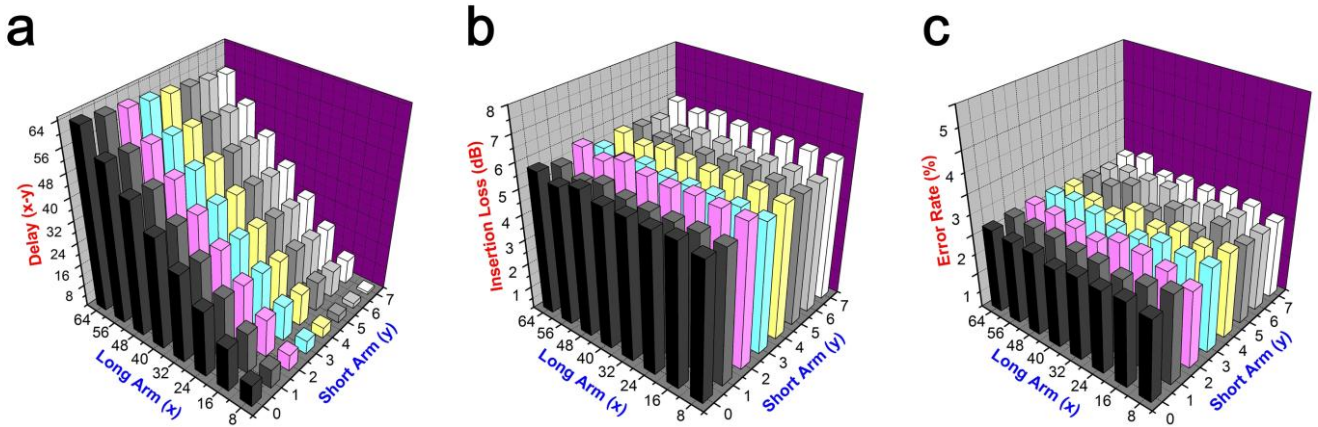

**FIG. 2.** Parameters of each delay of the variable interferometer. **a.** 64 variable delays of the interferometer, corresponding to 64 combinations ($x - y$) of the delay $x$ of the long arm and the delay $y$ of the short arm. For the 1-GHz, 1-64-bit variable-delay interferometer in our experiment, the dimensions of $x$, $y$, and $x - y$ are ns. **b.** IL of 64 delays of the variable interferometer. The average IL of the variable-delay interferometer is 5.60 dB. The maximum IL, 6.04 dB, occurs at a delay value of 46 ($x = 48, y = 2$), and the minimum IL, 5.30 dB, occurs at a delay value of 5 ($x = 8, y = 3$). **c.** Error rate of 64 delays. The average error rate is approximately 2.20% and was measured in the case in which the average photon number per pulse was 0.1 and transmission distance was 0 km. Each value of 64 delays was sorted out. The maximum value, 2.41%, occurred at a delay of 43 ($x = 48, y = 5$). The variance of these 64 error rates is approximately 0.011%.

The high precision of each delay, the good balance between the two arms, and the low insertion loss of the OS enable our variable interferometer scheme to exhibit very high scalability. Although the implemented variable interferometer is 1 GHz and 1-64 bits, we could expand it to 1 GHz and 1-1024 bits by using two $1 \times 32$ OSs. However, the change in the IL of the interferometer is only one round-trip IL per two $1 \times 2$ OSs, which is approximately 2.4 dB. Compared with three other feasible schemes (see Methods for details), the implemented Faraday-Michelson variable interferometer exhibits optimal performance. Compared with the passive scheme, the implemented variable interferometer only needs two detectors. Compared with the simple active scheme, the implemented variable interferometer has better IL balance between the two arms. Compared with the balanced active scheme based on the Mach-Zehnder interferometer structure, the implemented variable interferometer has almost the same IL but better stability. These advantages help the implemented variable interferometer obtain high visibility and good stability.

The photons from the variable-delay interferometer were detected by the gated SPD, which is based on InGaAs avalanche photodiodes and uses the sine-wave filtering method [23]. Two channels of this SPD were controlled independently to achieve approximately the same performance. The first channel was operated at a temperature of $-35°C$ with 20.23% detection efficiency and $0.85 \times 10^{-6}$ dark counts per gate; the second channel was operated at a temperature of $-50°C$ with 20.36% detection efficiency and $1.25 \times 10^{-6}$ dark counts per gate. The gate durations of the first and second channels were 154 and 178 ps, respectively, and the inherent dead time of both channels was 120 ns, which resulted in no more than one count during the detection of one packet. Except for dark counts, the afterpulse probability of SPD also contributes to the error rate of the system, and this contribution is independent of the fiber distance [24]. We developed one technique to effectively decrease the afterpulse probability and achieved afterpulse probability values of 1.25% and 0.95% for the two channels.

To estimate the error rate due to optical misalignments and the afterpulse effect, we set the average photon number per pulse to 0.1 and transmission distance to 0 km and then measured the average probability with which an erroneous detector clicked. The results for all 64 delays were sorted out, as shown in Fig. 2c. The average error rate is approximately 2.20%. The maximum value is 2.41% at a delay of 43 ($x = 48, y = 5$). The variance of these 64 error rates is approximately 0.011%.

Based on the above specific experimental parameters, the performance of our RRDPS QKD system was estimated, and the secret key rate was maximized by choosing the optimal average photon number per pulse $\mu$ for each fiber length (see Methods for details). The corresponding error rates and secret key rates were simulated after obtaining the optimal $\mu$ values. The simulation results regarding the secret key rate, error rate, and optimal $\mu$ are denoted by the solid lines in Figs. 3a, b, and c, respectively. Setting the real-life average photon number per pulse close to the optimal one, we measured the yield per packet Q and the bit error rate $e_b$ and then calculated the secret key rates R (using Eq.(3)) and $R_{ft}$ (using Eq. (4)) for six different fiber lengths: 20, 50, 70, 80, 85, and 90 km. The experimental results are denoted by the circles and triangles in Fig. 3. In Fig. 3a, the circles represent the key rate per pulse for the asymptotic case, and the triangles are for the finite-key size case, in which the security parameter is $3 \times 2^{-80}$ and the number of packet emitted $N_{em} = 2 \times 10^8$. Even at a transmission distance of 80 km, $R_{ft}$ is approximately 76% of R, which exhibits better tolerance of the finite-sized-key effect of RRDPS QKD. From Fig. 3a, it is clear that the secret key bits can be distributed over a 90-km transmission distance. From Fig. 3b, the increase in the error rate primarily comes from the contribution of the dark counts of the SPD. If a superconducting nanowire SPD (e.g., ID280 from ID Quantique) were used, then the secure transmission distance would increase substantially (e.g., 160 km using ID280).

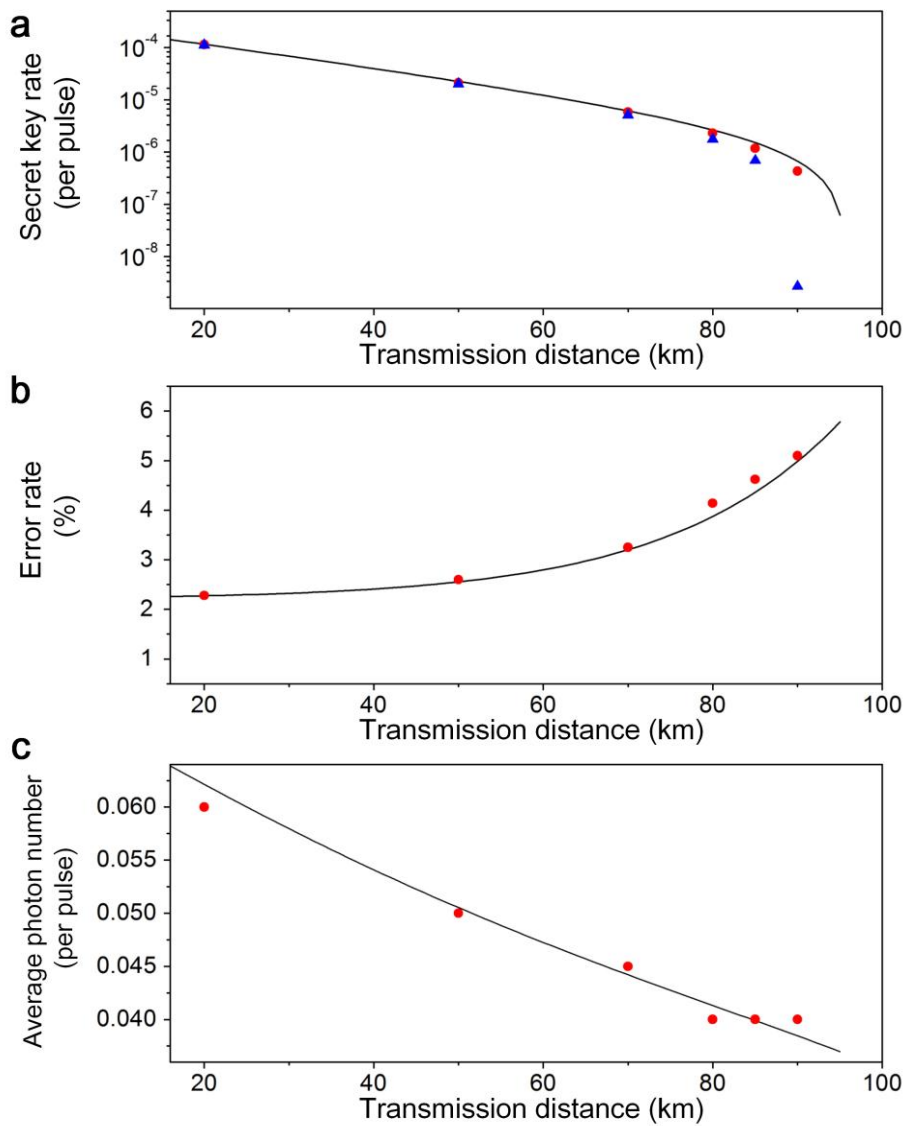

**FIG. 3.** Numerical simulation and experimental results. The solid curve indicates the simulation results, and the circles denote the experimental results for transmission distances of 20, 50, 70, 80, 85, and 90 km. **a.** Secret key rate (per pulse) v.s. transmission distance. The circles are for the asymptotic situation; the triangles are for the finite-key-size case, in which the security parameter was $3 \times 2^{-80}$ and the number of packets emitted was $N_{em} = 2 \times 10^8$. **b.** Error rate v.s. transmission distance. The solid curve denotes the bit error rate predicted by the numerical simulation with $e_s = 2.2\%$, where $e_s$ is the average probability that a signal hit the wrong detector. **c.** The average photon

number (per pulse) v.s. transmission distance. The solid curve denotes the optimal values calculated using the numerical simulation. According to these values, the experimental average photon numbers (per pulse) were 0.06 at 20 km, 0.05 at 50 km, 0.045 at 70 km, and 0.04 at 80, 85 and 90 km. It is clear that the secret key bits can be distributed over a 90-km transmission distance.

To summarize, we have successfully demonstrated an implementation of the RRDPS QKD protocol. As far as we know, this demonstration is the first active QKD experiment that did not require monitoring of the signal disturbance. Our achievement demonstrates that this novel protocol is feasible under practical conditions. The security and performance of the RRDPS protocol depend on the number of pulses L of each packet. In our implementation, L=65 is sufficiently large to demonstrate the potential advantages of this protocol. Although it seems to be a great challenge, four feasible schemes of the variable interferometer were developed to meet the requirement of large delay numbers. In addition to good stability, our implemented scheme exhibits very high scalability. Specifically, our variable-delay interferometer can be easily extended from 64 to 1024 bits, but the increase in the IL of the interferometer is only approximately 2.4 dB. A larger L will increase the key rate per packet and improve the bit error rate tolerance and channel loss. However, several issues must be addressed before realization of RRDPS QKD with larger L values. First, the stability of interferometer may be a challenge when L is very large because the phase would shift substantially, even if the environmental temperature changes little. To overcome this problem, finer active temperature control may be indispensable. Second, if L becomes larger, then the key rate per second may be decreased because more time would be spent to encode and detect each packet. To alleviate these effects and improve the performance of the RRDPS QKD, the repetition rate of a system with larger L must be increased. Despite these challenges, an RRDPS QKD system with large L and high repetition rate promises to be realized in the near future.

# Methods

**Schemes of the variable-delay interferometer** The variable-delay interferometer is the core part of Bob's setup. We propose four typical schemes to implement the variable-delay interferometer. Taking the 1-64-bit variable-delay interferometer as an example, three schemes are shown in Fig. 4. The passive scheme is based on a $1 \times 64$ BS, followed by 64 FMIs (see Fig. 4a), of which the delays are $\{1,2,\ldots,64\}$. Each interferometer uses one PZT cylinder to achieve better precision and compensate for the phase shift. All of the 64 interferometers are completely independent. The passive scheme has the advantage of high speed but requires too many SPDs. Fig. 4b and Fig. 4c show two active schemes based on the Mach-Zehnder interferometer structure. The simple scheme shown in Fig. 4b uses a $2 \times 2$ OS with $log_2(L-1)$ separate delays, and the balanced scheme shown in Fig. 4c uses a $1 \times N$ OS with $2N = 2\sqrt{(L-1)}$ separate delays. The basic unit in Fig. 4b is composed of a $2 \times 2$ OS and a PZT cylinder wrapped by an optical fiber with delay x. If the $2 \times 2$ OS is in the bar state, then the basic unit has delay 0; if the $2 \times 2$ OS is in the cross state, then the basic unit has delay x. There are three basic units in the short arm, with delays of $2^0 = 1$, $2^1 = 2$, and $2^2 = 4$. In the long arm, there are three basic units, with delays $2^3 = 8$, $2^4 = 16$, and $2^5 = 32$, and one PZT cylinder with delay $2^3 = 8$. By controlling the states of the basic unit, we can obtain delays of $\{0,1,2,3,4,5,6,7\}$ in the short arm and delays of $\{8,16,24,32,40,48,56,64\}$ in the long arm. Using this scheme, we only need to prepare $log_2 64 = 6$ types of sperate delays. However, the IL of the basic unit in the cross state is approximately twice the IL in the bar state. Therefore, the losses of the long arm and the short arm become unbalanced, e.g., for the variable-delay interferometer with delay 64, all of three basic units of the long arm are in the cross state, with $0.8 \times 2 \times 3 = 4.8$ dB, whereas all of three units of the short arm are in the bar state, with $0.8 \times 3 = 2.4$ dB, where 0.8 dB is the typical insertion loss of a $2 \times 2$ OS. This imbalance between the long and short arms makes the visibility of the variable-delay interferometer very poor for some delays. Fig. 4c is a balanced scheme to implement the variable-delay interferometer. In the short arm, $\{0,1,2,3,4,5,6,7\}$ delays lie between two $1 \times 8$ OSs, and in the long arm, $\{8,16,24,32,40,48,56,64\}$ delays are also located between two $1 \times 8$ OSs. To obtain a more stable interference effect, these two active schemes can also be implemented based on the FMI. However, the IL of the variable interferometers will increase, and the simple scheme in Fig. 4b will become more unbalanced. Based on the FMI, the final experimental scheme we chose is balanced and stable, but the IL is almost the same as that of the balanced active scheme shown in Fig. 4c.

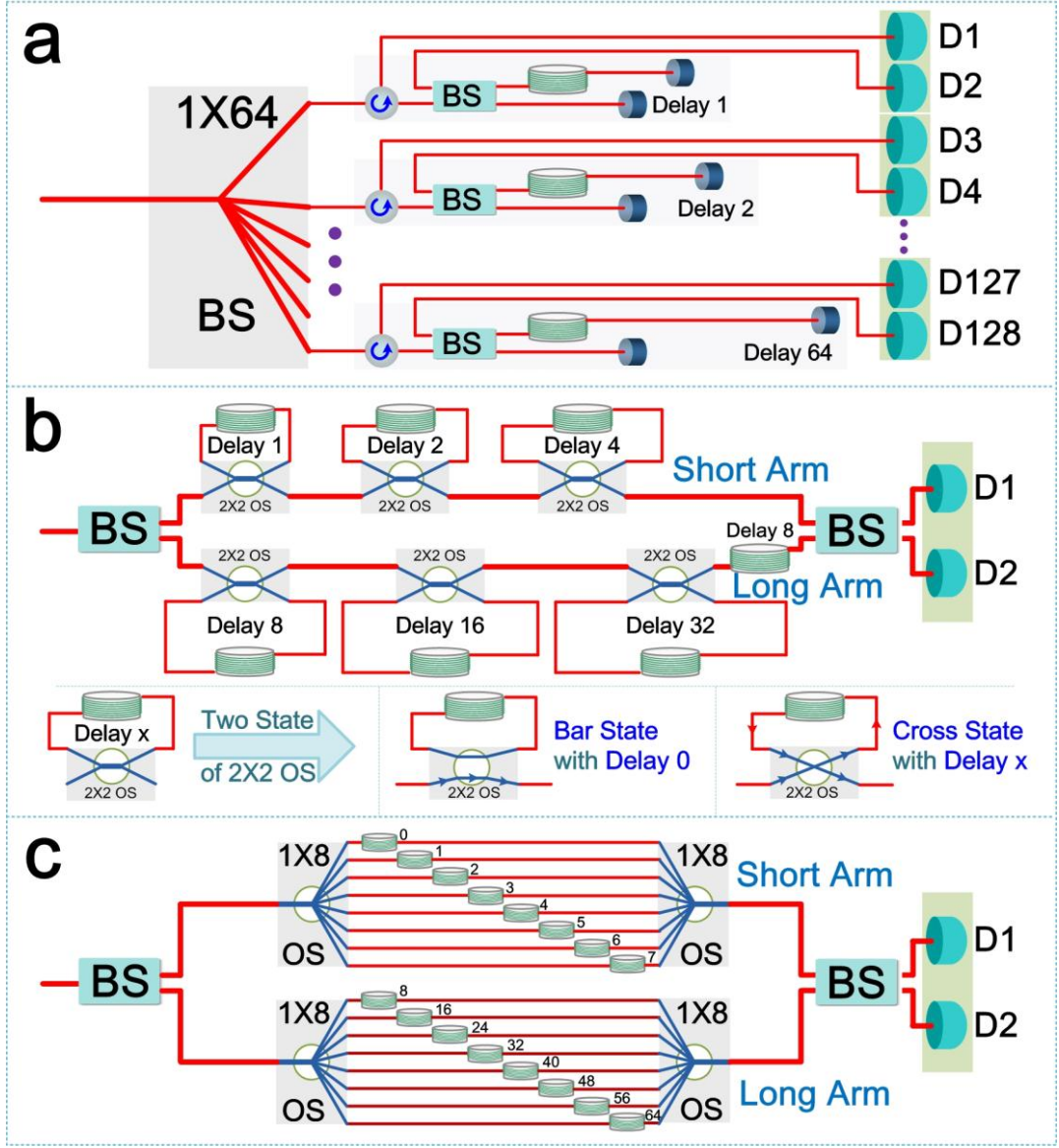

**FIG. 4.** Three schemes for the variable-delay interferometer. **a.** Passive scheme based on a $1 \times 64$ BS. **b.** Simple active scheme based on $2 \times 2$ OSs. **c.** Balanced active scheme based on $1 \times 8$ OSs.

**Performance estimate for the QKD system** To estimate the performance of our RRDPS QKD system and determine the optimal average photon number to be set, we used a numerical simulation based on parameters of the system. In the asymptotic case, the yield for an $i$-photon packet detected with an **r**-pulse delay interferometer is approximately

$$Y_{r,i} = \frac{L-r}{L}\left(1 - (1-\eta)^i + d\right),$$

where $\eta = \eta_D 10^{-(\alpha l + \alpha_{IL})/10}$, $\eta_D$ is the detection efficiency of the SPD, $\alpha$ and $l$ are the loss coefficient and the length of the fiber channel, respectively, $\alpha_{IL}$ is the IL of the variable interferometer, and $d$ is the total dark count rate of the two channels of the SPD. The average yield for an $i$-photon packet is given by

$$Y_i = \sum_{r=1}^{L-1} \frac{Y_{r,i}}{L-1} = \frac{1}{2}\left(1 - (1-\eta)^i + d\right).$$

Then, considering each packet is a train of coherent pulses with the average photon number of μ, we can give the average yield of each packet as

$$Q = \sum_{i=0}^{\infty} e^{-L\mu} \frac{(L\mu)^i}{i!} Y_i = \frac{1}{2}\left(1 - e^{-\eta L\mu} + d\right).$$

And the bit error rate $e_b$ is

$$e_b = \frac{e_s(1 - e^{-\eta\mu}) + d/2}{1 - e^{-\eta\mu} + d},$$

in which $e_s$ is the average probability that a signal hit the wrong detector. With Q and $e_b$, we are able to calculate the secret key rate R using Eq. (3). Thus, the optimal μ can be determined in a straightforward manner, and the corresponding Q, $e_b$, and R are determined through numerical methods. To maximize the secret key rate of our implementation, the parameters we used are as follows: the loss coefficient $\alpha = 0.2 dB/km$, the average IL of the 1-GHz, 1-64-bit variable-delay interferometer $\alpha_{IL}$ is 5.60 dB, the average detection efficiency $\eta_D$ and total dark count rate d of the two channels of the SPD are 20.3% and $2.1 \times 10^{-6}$, respectively, the error rate $e_s$ is 2.20%, and the efficiency of error correction f is 1.1.

# Acknowledgements

The authors thanks Jin-Dong Wang and Jing-Zheng Huang for scientific discussions. This work was supported by the National Natural Science Foundation of China (Grant Nos. 61475148, 61201239, 61205118, and 11304397), National Basic Research Program of China (Grants Nos. 2011CBA00200 and 2011CB921200), and the Strategic Priority Research Program (B) of the Chinese Academy of Sciences (Grant Nos. XDB01030100 and XDB01030300).

# Author contributions

The first two authors, Shuang Wang and Zhen-Qiang Yin, contributed equally to this work. Shuang Wang, Zhen-Qiang Yin, Wei Chen, Guang-Can Guo, and Zheng-Fu Han conceived and designed the experiments. Shuang Wang, Wei Chen, and Xiao-Tian Song designed the variable-delay interferometer. Shuang Wang, De-Yong He and Zheng Zhou designed the control and detection parts of the system. Zhen-Qiang Yin, Wei Chen and Li-Jun Zhang designed the software. All authors performed the experiments. Zhen-Qiang Yin and Hong-Wei Li performed the simulation and analyzed the data. Shuang Wang and Zhen-Qiang Yin wrote the paper.

# Corresponding authors

Wei Chen (weich@ustc.edu.cn) and Zheng-Fu Han (zfhan@ustc.edu.cn).